\documentclass[12pt]{article}

\usepackage[dvips,final]{graphicx}
\usepackage{graphicx}

\title{To the theory of helical waveguides}
\author{ A.S. Yurkov\thanks{fitec@mail.ru}}

\begin{document}

\maketitle

\begin{center}
Omsk scientific center SB RAS, 644024, K.Marksa 15, Omsk, Russia.
\end{center}

\abstract{It makes sense to consider a helical waveguide with a fine pitch approximately, replacing the turns with anisotropic conductivity: infinite along the turns and zero across them. This approach has been known for a long time, but calculation formulas within it have only been obtained for the case where the winding does not contain a dielectric core. This paper addresses this gap in the theory: calculation formulas are obtained for the case where the waveguide contains a dielectric with a certain permittivity and magnetic permeability. An equation determining the slowing factor is found, and a method for its numerical solution is proposed. Explicit formulas are obtained for the wave impedance.}

\section{Introduction}


Helical waveguides are structures important for practical applications. In particular, such waveguides are used as structural components of antennas (see, for example, \cite{5069229,aslan2023study}) and resonators (for example, \cite{826851,spencer1987review}). Furthermore, a single-layer inductor can be considered as a section of such a waveguide, which automatically accounts for possible parasitic resonances.

The general theory of spiral structures is quite complex due to the fact that in this case, the variables involved in Maxwell's equations are not separable \cite{pierce1947theory}. From a fundamental standpoint, separability of variables is not a necessary condition; Maxwell's equations can always be solved numerically on a grid, especially since many appropriate software packages are currently available. However, this approach requires extremely large computational resources, especially in situations where the spiral pitch is small and there are many turns.

At the same time, for the limiting case of a small helix pitch, when a direct numerical solution is problematic, an approximate approach \cite{corum2001rf} is known. This approach is based on replacing the helix turns with a surface with anisotropic conductivity: along the turns the conductivity is infinite, and across them it is zero. This approximation is valid if the wavelength (taking into account its slowing down in the waveguide) is much greater than the helix pitch. This approach, despite being known for a long time, is still actively cited in more modern works \cite{oh2013extremely,wang2022capacitively,martines2021helical}.

In the work cited above \cite{corum2001rf} assumed that the coil was surrounded by a medium with unit permittivity and permeability (essentially air). However, in reality, a dense coil of thin conductor is most often wound on a frame with a permittivity different from unity. It can also be filled with ferrite or another magnetic material. Under these conditions, theory \cite{corum2001rf} requires modification. The development of such a modified theory is the goal of this paper, and here we restrict ourselves to the case where the coil frame is solid.

\section{Representation of a spiral waveguide wave as a superposition of $H$- and $E$-waves}

We consider a circular dielectric rod made of a material with relative permittivity $\varepsilon$ and magnetic permeability $\mu$. The radius of the rod is denoted by $R$. The rod is wound with a single-layer winding with a pitch $h$, the thickness of which we neglect (accounting for the thickness is in any case impossible within the framework of our approximation). Outside the rod there is a medium with properties close to a vacuum (for example, air), so outside the rod $\varepsilon=\mu=1$. Cartesian coordinate axis $z$ goes along the rod axis,  $x$ and $y$ axes directed arbitrary. We will also use a cylindrical coordinate system $z,\alpha,\rho$, where $\rho$ is the radial coordinate, $\alpha$ is the polar angle.

It is well known that if an electromagnetic wave is proportional to $e^{-jkz}$ along one of the spatial coordinates, in our case $z$ (the waveguide case), then the transverse components of the field are reflected through the longitudinal ones. The only exception to this rule are purely transverse $T$-waves, for which the longitudinal components are zero. Obviously, the case of a $T$-wave does not apply to a helical waveguide, since the turns along which the electric field is zero are inclined with respect to the waveguide axis. From this, we can conclude that in the case considered here, the waves are either $E$-waves, or $H$-waves, or a superposition of these two types of waves. It will become clear below that it is precisely the third, and only the third case  is realized.

In the case of time dependence of the form $\sim e^{j\omega t}$, the longitudinal components of the waves in the waveguide satisfy  two-dimensional homogeneous Helmholtz equations:
\begin{equation}
\left(\frac{\partial^2}{\partial x^2} + \frac{\partial^2}{\partial y^2}\right)E_z + 
(\varepsilon\mu k_0^2 - k^2)E_z = 0 \, ,
\end{equation}
\begin{equation}
\left(\frac{\partial^2}{\partial x^2} + \frac{\partial^2}{\partial y^2}\right)H_z + 
(\varepsilon\mu k_0^2 - k^2)H_z = 0 \, ,
\end{equation}
\begin{equation}
k_0^2=\varepsilon_0\mu_0\omega^2 \, ,
\end{equation}
where $\varepsilon_0$ and $\mu_0$ are the electric and magnetic constants, respectively. In addition, outside the waveguide, we must substitute $\varepsilon=\mu=1$. In the $E$-wave, $H_z=0$ (the trivial solution of any linear homogeneous equation), and in the $H$-wave, $E_z=0$. In the general case, which will be realized in a spiral waveguide, $E_z\ne 0 \ne H_z$.

In cylindrical coordinates the equations written above have the following form:
\begin{equation}
\frac{1}{\rho}\frac{\partial}{\partial\rho}\left(\rho\frac{\partial E_z}{\partial\rho}\right) +
\frac{1}{\rho^2}\frac{\partial^2 E_z}{\partial\alpha^2} +
(\varepsilon\mu k_0^2 - k^2)E_z = 0 \, ,
\end{equation}
\begin{equation}
\frac{1}{\rho}\frac{\partial}{\partial\rho}\left(\rho\frac{\partial H_z}{\partial\rho}\right) +
\frac{1}{\rho^2}\frac{\partial^2 H_z}{\partial\alpha^2} +
(\varepsilon\mu k_0^2 - k^2)H_z = 0 \, .
\end{equation}
In these equations the variables are separated, and the general solution can be written immediately:
\begin{equation}
E_z = C_E e^{jn\alpha} Z_n(\gamma\rho) \, ,
\end{equation}
\begin{equation}
H_z = C_H e^{jn\alpha} Z_n(\gamma\rho) \, ,
\end{equation}
where $C_E$ and $C_H$ are constants to be determined later, $\gamma=\sqrt{|\varepsilon\mu k_0^2 - k^2|}$, $Z_n$ is one or another cylindrical function of the $n$-th order, either ordinary or modified depending on the sign of $\varepsilon\mu k_0^2 - k^2$ (modified if the sign is negative). Here we restrict ourselves to the fundamental azimuthal mode, which has a zero cutoff frequency, which is most interesting for practice (and for sufficiently low frequencies, which is the only one possible). Therefore, in what follows, $n=0$.

Outside the waveguide, i.e., for $\rho > R$, the wave should attenuate with increasing radial coordinate. Only in this case is a channeled wave obtained. Therefore, $k_0^2 - k^2 < 0$ (i.e., outside the waveguide, the wave should be slow), and outside the waveguide the MacDonald function $K_0(\rho\sqrt{k^2 - k_0^2})$ should be taken as the cylindrical function. It cannot be ruled out a priori that, unlike the case of an unfilled waveguide, inside, depending on the dielectric parameters, not only a slow but also a fast wave may be produced (fast relative to free waves in such a dielectric, but still slow relative to waves in free space). Accordingly, as a cylindrical function we must take either the modified Bessel function $I_0(\rho\sqrt{k^2 - \varepsilon\mu k_0^2})$, or the ordinary Bessel function $J_0(\rho\sqrt{ \varepsilon\mu k_0^2-k^2})$. In each of these cases, the value under the square root must be positive.

To summarize, we can write the following representations for the longitudinal components of the field in our case. For the case of a slow wave inside a waveguide:
\begin{equation}
E_z(\rho > R) = C_{E+}K_0\left(\rho\sqrt{k^2-k_0^2}\right)\, , 
\qquad E_z(\rho < R) = C_{E-} I_0\left(\rho\sqrt{k^2-\varepsilon\mu k_0^2}\right) \, ,
\end{equation}
\begin{equation}
H_z(\rho > R) = C_{H+}K_0\left(\rho\sqrt{k^2-k_0^2}\right))\, , 
\qquad H_z(\rho < R) = C_{H-} I_0\left(\rho\sqrt{k^2-\varepsilon\mu k_0^2}\right) \, .
\end{equation}
For the case of a fast wave inside a waveguide:
\begin{equation}
E_z(\rho > R) = C_{E+}K_0\left(\rho\sqrt{k^2-k_0^2}\right)\, , 
\qquad E_z(\rho < R) = C_{E-} J_0\left(\rho\sqrt{\varepsilon\mu k_0^2-k^2}\right) \, ,
\end{equation}
\begin{equation}
H_z(\rho > R) = C_{H+}K_0\left(\rho\sqrt{k^2-k_0^2}\right))\, , 
\qquad H_z(\rho < R) = C_{H-} J_0\left(\rho\sqrt{\varepsilon\mu k_0^2-k^2}\right)) \, .
\end{equation}
Here the constants $C$ are provided with an additional index + (field outside the waveguide) or - (inside the waveguide).

\section{Dispersion law}

First, we need to determine the dispersion law, that is the dependence of $k$ on $\omega$. As stated above, the field along the turns must be zero (both outside and inside the winding). From simple geometric considerations it follows that the field along a turn has two components: a contribution from $E_\alpha$ and a contribution from $E_z$. The first contribution is proportional to the cosine of the helix angle $\phi$, the second to the sine. Thus,
\begin{equation}
\label{Eq:Et}
\left. \cos\phi E_\alpha + \sin\phi E_z \right|_{\rho=R} = 0 \, .
\end{equation}
Or
\begin{equation}
\label{Eq:Ettg}
\left.  E_\alpha + \tan\phi E_z \right|_{\rho=R} = 0 \, .
\end{equation}
Note that these are actually two expressions: it can be applied to both the field outside the winding and the field inside the winding. The slope tangent is easily determined from the same geometric considerations:
\begin{equation}
\label{Eq:tg}
\tan\phi = \frac{h}{2\pi R} \, .
\end{equation}

In total, we have four constants: two amplitudes (for the E- and H-waves) outside and inside the helix. Two more equations are needed; they are obtained from the continuity condition of the electric field component perpendicular to the turns and the continuity condition of the magnetic field component tangential to the turns (the perpendicular component of the magnetic field experiences a jump caused by the current flowing through the winding). Introducing the notation $R_+$ for the limiting value $\rho\to R$ outside, and $R_-$ inside, we obtain the following:
\begin{equation}
\label{Eq:En}
\left.  -\sin\phi E_\alpha + \cos\phi E_z \right|_{\rho=R_+} = 
\left.  -\sin\phi E_\alpha + \cos\phi E_z \right|_{\rho=R_-} \, . 
\end{equation} 
Taking into account (\ref{Eq:Ettg}) this equation gives the continuity of the component $E_z$ at $\rho=R$:
\begin{equation}
\label{Eq:Ezcont}
  \left. E_z \right|_{\rho=R_+} =   \left. E_z \right|_{\rho=R_-} \, , 
\end{equation} 
This condition can  replace (\ref{Eq:En}). The continuity equation for the tangential component of the magnetic field has the following form:
\begin{equation}
\label{Eq:Ht}
\left.  \cos\phi H_\alpha + \sin\phi H_z \right|_{\rho=R_+} = 
\left.  \cos\phi H_\alpha + \sin\phi H_z \right|_{\rho=R_-} \, , 
\end{equation} 
It is also convenient to rewrite this equation in terms of the tangent defined by the equation (\ref{Eq:tg}):
\begin{equation}
\label{Eq:Htt}
\left.  H_\alpha + \tan\phi H_z \right|_{\rho=R_+} = 
\left.  H_\alpha + \tan\phi H_z \right|_{\rho=R_-} \, . 
\end{equation} 
The formulas (\ref{Eq:Et}), (\ref{Eq:En}) and (\ref{Eq:Ht}) are easy to understand if we notice that the matrix
\begin{equation}
\left(
\begin{array}{cc}
\cos\phi & \sin\phi \\
-\sin\phi & \cos\phi
\end{array}
\right) 
\end{equation}
is nothing but the rotation matrix by an angle $\phi$ in two-dimensional space tangent to the winding.

Now we need to express the components of the fields $E_\alpha$, $H_\alpha$, $E_\rho$ and $H_\rho$ in terms of $E_z$ and $H_z$. Using Maxwell's equations, we obtain
\begin{equation}
\label{Eq:HalphaEz}
H_\alpha = \frac{j\omega\varepsilon\varepsilon_0}{k^2 -\varepsilon\mu k_0^2}
\frac{\partial E_z}{\partial\rho} \, ,
\end{equation}
\begin{equation}
\label{eq:EalphaHz}
E_\alpha = -\frac{j\omega\mu\mu_0}{k^2 -\varepsilon\mu k_0^2}\frac{\partial H_z}{\partial\rho} \, ,
\end{equation}
\begin{equation}
\label{eq:HrhoEalpha}
H_\rho = - \frac{k}{\omega\mu\mu_0} E_\alpha    \, ,
\end{equation}
\begin{equation}
\label{eq:ErhoHalpha}
E_\rho = \frac{k}{\omega \varepsilon\varepsilon_0} H_\alpha  \, .
\end{equation}
Using these equalities and substituting the above representation of the field components through cylindrical functions, we can write the boundary conditions as a system of linear algebraic equations (SLAE) for the constants $C_{E+}$, $C_{E_-}$, $C_{H+}$, and $C_{H-}$. For the case of slow waves inside the waveguide, we obtain the following:
\begin{equation}
\label{EQ:SysI0}
\left\{
\begin{array}{l}
\displaystyle
\frac{j\omega\mu_0}{\sqrt{k^2 - k_0^2}}
C_{H+}K_1\left(R\sqrt{k^2 - k_0^2}\right)
 + \tan\phi C_{E+} K_0\left(R\sqrt{k^2 - k_0^2}\right)  = 0 \, , \\ \\
 \displaystyle
-\frac{j\omega\mu\mu_0}{\sqrt{k^2 - \varepsilon\mu k_0^2}}
C_{H-}I_1\left(R\sqrt{k^2 - \varepsilon\mu k_0^2}\right)
 + \tan\phi C_{E-} I_0\left(R\sqrt{k^2 - \varepsilon\mu k_0^2}\right)  = 0 \, ,\\ \\
\displaystyle
C_{E+}K_0\left( R\sqrt{k^2-k_0^2}\right) -  C_{E-}I_0\left( R\sqrt{k^2-\varepsilon\mu k_0^2}\right) = 0 \, , \\ \\ 
\displaystyle
\frac{j\omega\varepsilon_0}{\sqrt{k^2- k_0^2}}C_{E+}K_1\left(R\sqrt{k^2-k_0^2}\right)
- \tan\phi C_{H+}K_0\left(R\sqrt{k^2-k_0^2}\right)  + \\ \\
\displaystyle +
 \frac{j\omega\varepsilon\varepsilon_0}{\sqrt{k^2-\varepsilon\mu k_0^2}} 
C_{E-}I_1\left(R\sqrt{k^2-\varepsilon\mu k_0^2}\right) + 
\tan\phi C_{H-}I_0\left(R\sqrt{k^2-\varepsilon\mu k_0^2}\right) = 0 \, . 
\end{array}
\right.
\end{equation}
This is a homogeneous SLAE. Its solvability condition is the equality to zero of the corresponding determinant, which yields an equation defining $k$ as a function of $k_0$. In fact, this is the dispersion equation, defining the dependence of $k$ on $\omega$ (an implicit dependence, through the quantity $k_0$ proportional to $\omega$). The dispersion equation can also be obtained in a different equivalent way by successively expressing some quantities of $C$ in terms of others. In this particular case, the second method is more convenient due to the special form of the SLAE (\ref{EQ:SysI0}).

Before writing the dispersion equation, it is convenient to introduce new variables: 
\begin{equation}
\xi=\frac{k}{k_0} \, ,
\end{equation}
\begin{equation}
W_0=\sqrt{\frac{\mu_0}{\varepsilon_0}} \, ,
\end{equation}
\begin{equation}
A= Rk_0 =\frac{2\pi R}{\lambda} \, ,
\end{equation}
where $\lambda$ is the wavelength of a free electromagnetic wave with angular frequency $\omega$, propagating in free space. The quantity $W_0$ is the wave impedance of free space, the dimensionless quantity $\xi$ indicates how many times the phase velocity of the wave in the waveguide is less than the wave velocity in free space, and the meaning of $A$ is obvious from the formula. In this notation, the SLAE (\ref{EQ:SysI0}) takes the following form:
\begin{equation}
\label{Eq:sys1}
\left\{
\begin{array}{l}
\displaystyle
\frac{jW_0}{\sqrt{\xi^2 - 1}}
C_{H+}K_1\left(A\sqrt{\xi^2 - 1}\right)
 + \tan\phi C_{E+} K_0\left(A\sqrt{\xi^2 - 1}\right)  = 0 \, , \\ \\
 \displaystyle
-\frac{jW_0\mu}{\sqrt{\xi^2 - \varepsilon\mu}}
C_{H-}I_1\left(A\sqrt{\xi^2 - \varepsilon\mu }\right)
 + \tan\phi C_{E-} I_0\left(A\sqrt{\xi^2 - \varepsilon\mu}\right)  = 0 \, , \\ \\
\displaystyle
C_{E+}K_0\left( A\sqrt{\xi^2-1}\right) -  C_{E-}I_0\left( A\sqrt{\xi^2-\varepsilon\mu }\right) 
= 0 \, , \\ \\ 
\displaystyle
\frac{j}{W_0\sqrt{\xi^2-1}}C_{E+}K_1\left(A\sqrt{\xi^2-1}\right)
- \tan\phi C_{H+}K_0\left(A\sqrt{\xi^2-1}\right)  + \\ \\
\displaystyle +
 \frac{j\varepsilon}{W_0\sqrt{\xi^2-\varepsilon\mu}} 
C_{E-}I_1\left(A\sqrt{\xi^2-\varepsilon\mu}\right) + 
\tan\phi C_{H-}I_0\left(A\sqrt{\xi^2-\varepsilon\mu}\right) = 0 \, . 
\end{array}
\right.
\end{equation}
By successively eliminating various $C$ in this SLAE, we obtain three formulas that will be needed further,
\begin{equation}
\label{Eq:CE-viaCE+}
C_{E-} = C_{E+}\frac{K_0\left( A\sqrt{\xi^2-1}\right)}{I_0\left( A\sqrt{\xi^2-\varepsilon\mu }\right)} \, ,
\end{equation}
\begin{equation}
\label{Eq:CH-viaCH+}
C_{H-} = - C_{H+} \frac{\sqrt{\xi^2 - \varepsilon\mu}}{\mu\sqrt{\xi^2 - 1}}\cdot
\frac{K_1\left(A\sqrt{\xi^2 - 1}\right)}{I_1\left(A\sqrt{\xi^2 - \varepsilon\mu }\right)} \, ,
\end{equation}
\begin{equation}
\label{Eq:CE+viaCH+}
C_{E+}= - C_{H+} \frac{jW_0}{\tan\phi\sqrt{\xi^2 - 1}}\cdot
\frac{K_1\left(A\sqrt{\xi^2 - 1}\right)}{K_0\left(A\sqrt{\xi^2 - 1}\right)} \, ,
\end{equation}
and the solvability condition of the SLAE, which determines the dispersion law,
\begin{equation}
\label{Eq:Sys1m5}
\begin{array}{l}
\displaystyle
\mu \sqrt{\xi^2-\varepsilon\mu} K_1^2\left(A\sqrt{\xi^2-1}\right)
I_0\left(A\sqrt{\xi^2 - \varepsilon\mu }\right)I_1\left(A\sqrt{\xi^2 - \varepsilon\mu }\right) -  \\ \\
\displaystyle -
\mu \tan^2\phi(\xi^2 - 1)\sqrt{\xi^2-\varepsilon\mu}
K_0^2\left(A\sqrt{\xi^2 - 1}\right)
I_0\left(A\sqrt{\xi^2 - \varepsilon\mu }\right)I_1\left(A\sqrt{\xi^2 - \varepsilon\mu }\right)  + \\ \\
\displaystyle +
\varepsilon\mu\sqrt{\xi^2-1} 
I_1^2\left(A\sqrt{\xi^2-\varepsilon\mu}\right)K_0\left( A\sqrt{\xi^2-1}\right) 
K_1\left(A\sqrt{\xi^2 - 1}\right) -  \\ \\
\displaystyle - 
\tan^2\phi 
\cdot(\xi^2 - \varepsilon\mu)\sqrt{\xi^2-1}\cdot
I_0^2\left(A\sqrt{\xi^2-\varepsilon\mu}\right)K_0\left(A\sqrt{\xi^2 - 1}\right)
K_1\left(A\sqrt{\xi^2 - 1}\right) = 0 \, . 
\end{array}
\end{equation}
Note that in the particular case $\varepsilon=\mu=1$ this equation is radically simplified and can be reduced to a form that completely coincides with the corresponding equation from \cite{corum2001rf}. In the general case, however, it is not possible to simplify this equation.

Equation (\ref{Eq:Sys1m5}) is a transcendental equation, and its analytical solution is impossible. Apparently, the most suitable numerical method for solving this equation is the bisection method.
The result of such a solution for certain parameter values is shown in Fig.\ref{fig1}. It should be noted that all values in the graphs lie within the applicability range of the approximate theory under consideration. Indeed, for the theory to be applicable, the wavelength, taking into account the deceleration, must be much greater than the winding pitch. The maximum deceleration factor $\xi$ is approximately 160, and the minimum wavelength in free space is 2 meters. Moreover, the wavelength, taking into account the deceleration, is approximately 12 mm, which is much greater than the winding pitch of 1 mm.

\begin{figure}[h]
\begin{center}
\vspace{1cm}
\includegraphics[width=14cm, keepaspectratio]{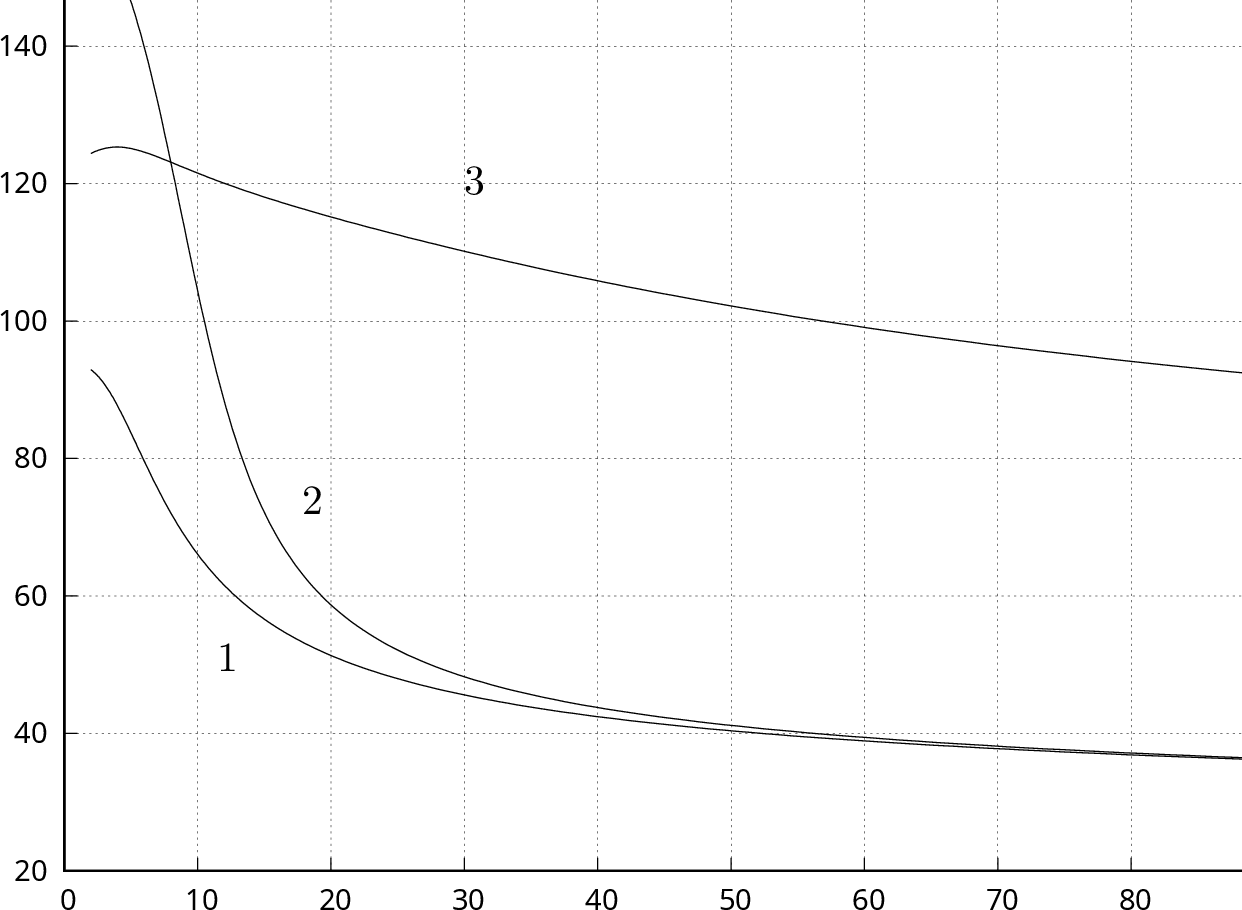}
\end{center}
\caption{The slowdown coefficient $\xi$ depending on the wavelength $\lambda$ for a waveguide with a diameter of 30 mm, with a winding pitch of 1 mm. 1: $\varepsilon=\mu=1$; 2: $\varepsilon=5$, $\mu=1$; 3: $\varepsilon=1$, $\mu=5$.}
\label{fig1}
\end{figure}

To obtain the dispersion equation for the case of fast waves inside a waveguide, there is no need to repeat all the calculations. It is sufficient to make the substitutions in (\ref{Eq:Sys1m5})
\begin{equation}
\begin{array}{l}
\displaystyle
I_0\left(A\sqrt{\xi^2 - \varepsilon\mu }\right) \to J_0\left(A\sqrt{\varepsilon\mu - \xi^2}\right) \, , \\ \\
\displaystyle
\sqrt{\xi^2 - \varepsilon\mu }\, I_1\left(A\sqrt{\xi^2 - \varepsilon\mu }\right) 
\to -\sqrt{\varepsilon\mu - \xi^2}\, J_1\left(A\sqrt{\varepsilon\mu - \xi^2}\right) \, .
\end{array}
\end{equation}
The result is the following:
\begin{equation}
\label{Eq:DispJ}
\begin{array}{l}
\displaystyle
\mu \sqrt{\varepsilon\mu-\xi^2} K_1^2\left(A\sqrt{\xi^2-1}\right)
J_0\left(A\sqrt{ \varepsilon\mu - \xi^2}\right)J_1\left(A\sqrt{ \varepsilon\mu - \xi^2}\right) -  \\ \\
\displaystyle -
\mu \tan^2\phi(\xi^2 - 1)\sqrt{\varepsilon\mu - \xi^2}
K_0^2\left(A\sqrt{\xi^2 - 1}\right)
J_0\left(A\sqrt{\varepsilon\mu - \xi^2}\right)J_1\left(A\sqrt{\varepsilon\mu - \xi^2}\right)  - \\ \\
\displaystyle -
\varepsilon\mu\sqrt{\xi^2-1} 
J_1^2\left(A\sqrt{\varepsilon\mu-\xi^2}\right)K_0\left( A\sqrt{\xi^2-1}\right) 
K_1\left(A\sqrt{\xi^2 - 1}\right) +  \\ \\
\displaystyle + 
\tan^2\phi 
\cdot( \varepsilon\mu - \xi^2)\sqrt{\xi^2-1}\cdot
J_0^2\left(A\sqrt{\varepsilon\mu-\xi^2}\right)K_0\left(A\sqrt{\xi^2 - 1}\right)
K_1\left(A\sqrt{\xi^2 - 1}\right) = 0 \, . 
\end{array}
\end{equation}
Attempts to solve (\ref{Eq:DispJ}) using the bisection method showed that, for the parameters shown in Fig.\ref{fig1}, real solution to this equation appears only for very large values of $\mu$ and/or $\varepsilon$. Such solutions are unlikely to be of significant practical interest. Therefore, we will not further consider the possibility of fast waves inside the waveguide, but if necessary, their dispersion law can easily be found using the formulas written above.

\section{Electromagnetic field expressed in terms of the current flowing through the winding}

Proceeding in the same way as in the elementary theory of the solenoid, we can write an equation that relates the jump in the z-component of the magnetic field and the current I flowing through the winding:
\begin{equation}
\left. H_z\right|_{\rho=R_+} - \left. H_z\right|_{\rho=R_-} =
C_{H_+}K_0\left(A\sqrt{\xi^2-1}\right) - C_{H_-}I_0\left(A\sqrt{\xi^2-\varepsilon\mu}\right) 
 =  \frac{I}{h} \, .
\end{equation}
Combining this equality with the formulas (\ref{Eq:CE-viaCE+}), ({\ref{Eq:CH-viaCH+}), (\ref{Eq:CE+viaCH+}), we obtain a SLAE from which all four constants of $C$ can be expressed through $I$. As a result, we obtain the following representations for the longitudinal components of the field:
\begin{equation}
H_z(x > 1)  =
\frac{I}{h}\cdot  \frac{ \mu\sqrt{\xi^2-1} I_1\left(A \sqrt{\xi^2-\varepsilon\mu} \right) }
{ B }\cdot K_0\left(x A \sqrt{\xi^2-1} \right) \, ,
\end{equation}
\vspace{5mm}
\begin{equation}
H_z(x < 1)  = - \frac{I}{h}\cdot
\frac{ \sqrt{\xi^2-\varepsilon\mu} K_1\left(A \sqrt{\xi^2-1} \right) }
{B}\cdot I_0\left(xA\sqrt{\xi^2-\varepsilon\mu} \right)
\, ,
\end{equation}
\vspace{5mm}
\begin{equation}
E_z(x>1) =  - \frac{I}{h}\cdot\frac{j \mu W_0
K_1\left(A \sqrt{\xi^2-1} \right) 
I_1\left(A \sqrt{\xi^2-\varepsilon\mu} \right) }
{\tan\phi K_0\left(A \sqrt{\xi^2-1} \right)B}
\cdot K_0\left(xA\sqrt{\xi^2-1} \right)  \, ,
\end{equation}
\vspace{5mm}
\begin{equation}
E_z(x<1) =  - \frac{I}{h}\cdot\frac{j \mu W_0
K_1\left(A \sqrt{\xi^2-1} \right) 
I_1\left(A \sqrt{\xi^2-\varepsilon\mu} \right) }
{\tan\phi I_0\left(A \sqrt{\xi^2-\varepsilon\mu} \right)B}
\cdot I_0\left(xA\sqrt{\xi^2-\varepsilon\mu} \right)  \, ,
\end{equation}
where
\begin{equation}
\label{eq:B}
B =  \mu\sqrt{\xi^2-1} I_1\left(A\sqrt{\xi^2-\varepsilon\mu}\right) K_0\left(A\sqrt{\xi^2-1} \right) +
\sqrt{\xi^2-\varepsilon\mu} I_0\left(A\sqrt{\xi^2-\varepsilon\mu}\right) K_1\left(A\sqrt{\xi^2-1} \right) \, ,
\end{equation}
and the normalized radial coordinate was introduced:
\begin{equation}
x = \frac{\rho}{R} \, .
\end{equation}
Then it only remains to use the formulas (\ref{Eq:HalphaEz}), (\ref{eq:EalphaHz}), (\ref{eq:HrhoEalpha}), (\ref{eq:ErhoHalpha}) and, taking into account that $\partial/\partial\rho =R^{-1}\partial/\partial x$, after some algebraic transformations, obtain a representation for the transverse components of the field:
\begin{equation}
\label{eq:Halph>}
H_\alpha(x>1) = - \frac{I}{h}\cdot\frac
{\mu I_1\left(A\sqrt{\xi^2-\varepsilon\mu}\right)K_1\left(A\sqrt{\xi^2-1}\right)}
{\sqrt{\xi^2-1}\tan\phi K_0\left(A\sqrt{\xi^2-1}\right)B}K_1\left(xA\sqrt{\xi^2-1}\right) \, ,
\end{equation}
\begin{equation}
\label{eq:Halph<}
H_\alpha(x<1) =  \frac{I}{h}\cdot\frac
{\varepsilon\mu I_1\left(A\sqrt{\xi^2-\varepsilon\mu}\right)K_1\left(A\sqrt{\xi^2-1}\right)}
{\sqrt{\xi^2-\varepsilon\mu}\tan\phi I_0\left(A\sqrt{\xi^2-\varepsilon\mu}\right)B}
I_1\left(xA\sqrt{\xi^2-\varepsilon\mu}\right) \, ,
\end{equation}
\begin{equation}
\label{eq:Ealph>}
E_\alpha(x>1) = \frac{I}{h}\cdot
\frac{jW_0\mu I_1\left(A\sqrt{\xi^2-\varepsilon\mu}\right)}{B}K_1\left(xA\sqrt{\xi^2-1}\right) \, ,
\end{equation}
\begin{equation}
\label{eq:Ealph<}
E_\alpha(x<1) = \frac{I}{h}\cdot
\frac{jW_0\mu K_1\left(A\sqrt{\xi^2-1}\right)}{B}I_1\left(xA\sqrt{\xi^2-\varepsilon\mu}\right) \, ,
\end{equation}
\begin{equation}
\label{eq:Hrho>}
H_\rho(x>1) = - \frac{I}{h}\cdot
\frac{j\mu\xi I_1\left(A\sqrt{\xi^2-\varepsilon\mu}\right)}{B}K_1\left(xA\sqrt{\xi^2-1}\right) \, ,
\end{equation}
\begin{equation}
\label{eq:Hrho<}
H_\rho(x<1) = - \frac{I}{h}\cdot
\frac{j\xi K_1\left(A\sqrt{\xi^2-1}\right)}{B}I_1\left(xA\sqrt{\xi^2-\varepsilon\mu}\right) \, ,
\end{equation}
\begin{equation}
\label{eq:Erho>}
E_\rho(x>1) = - \frac{I}{h}\cdot\frac
{\mu\xi W_0 I_1\left(A\sqrt{\xi^2-\varepsilon\mu}\right)K_1\left(A\sqrt{\xi^2-1}\right)}
{\sqrt{\xi^2-1}\tan\phi K_0\left(A\sqrt{\xi^2-1}\right)B}K_1\left(xA\sqrt{\xi^2-1}\right) \, ,
\end{equation}
\begin{equation}
\label{eq:Erho<}
E_\rho(x<1) = \frac{I}{h}\cdot\frac
{\mu\xi W_0 I_1\left(A\sqrt{\xi^2-\varepsilon\mu}\right)K_1\left(A\sqrt{\xi^2-1}\right)}
{\sqrt{\xi^2-\varepsilon\mu}\tan\phi I_0\left(A\sqrt{\xi^2-\varepsilon\mu}\right)B}
I_1\left(xA\sqrt{\xi^2-1}\right) \, .
\end{equation}
%


Note that, according to the given formulas, the components $E_z$, $E_\alpha$, and $B_\rho=\mu\mu_0 H_\rho$ on the winding (at $x=1$) are continuous, while $E_\rho$, $H_z$, and $H_\alpha$ experience a jump. This is as it should be, since the tangential components of the magnetic field experience a jump due to the presence of the tangential current, and the normal component of the electric field experiences a jump due to the presence of charges, the existence of which at a current $\sim e^{jkz}$ is inevitable due to charge conservation. From the given formulas, the continuity of $E_z$, $E_\alpha$, and $B_\rho$ is immediately obvious. It is also immediately obvious that the jump in $H_z$ has the correct value. In order to ensure the correct value of the jumps $H_\alpha$ and $D_\rho=\varepsilon\varepsilon_0E_\rho$, it is necessary to perform some algebraic transformations taking into account the formulas (\ref{Eq:Sys1m5}) and (\ref{eq:B}).

\section{Wave power, wave impedance and equivalent long line}

Using the formulas (\ref{eq:Halph>}), (\ref{eq:Halph<}), (\ref{eq:Hrho>}), (\ref{eq:Hrho<}), (\ref{eq:Ealph>}), (\ref{eq:Ealph<}), (\ref{eq:Erho>}), (\ref{eq:Erho<}), we can write the following expressions for the $z$-component $\Pi_z$ of the Poynting vector:
\begin{equation}
\Pi_z = {\rm Re} \, \frac{1}{2}\left( H^*_\alpha E_\rho - H^*_\rho E_\alpha\right) \, ,
\end{equation}
\begin{equation}
\label{eq:Piext}
\begin{array}{l}
\displaystyle
\Pi_z (x>1) = \frac{I^2\mu^2\xi W_0}{2h^2B^2}
I^2_1\left(A\sqrt{\xi^2-\varepsilon\mu}\right) \times \\ \\
\displaystyle \times
\left( \frac
{K^2_1\left(A\sqrt{\xi^2-1}\right)}
{(\xi^2-1)\tan^2\phi K^2_0\left(A\sqrt{\xi^2-1}\right)} + 1
 \right) 
K^2_1\left(xA\sqrt{\xi^2-1}\right) 
 \, ,
\end{array}
\end{equation}
\begin{equation}
\label{eq:Pint}
\begin{array}{l}
\displaystyle
\Pi_z (x<1) = \frac{I^2\mu\xi\ W_0}{2h^2 B^2}\cdot K^2_1\left(A\sqrt{\xi^2-1}\right) \times \\ \\
\displaystyle \times
\left(\frac {\varepsilon\mu I^2_1\left(A\sqrt{\xi^2-\varepsilon\mu}\right)}
{(\xi^2-\varepsilon\mu)\tan^2\phi I^2_0\left(A\sqrt{\xi^2-\varepsilon\mu}\right)}
 +  1 \right)
I^2_1\left(xA\sqrt{\xi^2-\varepsilon\mu}\right)
 \, .
\end{array}
\end{equation}

The power $P$ carried by the wave is the integral of $\Pi_z$ over $2\pi\rho d \rho = 2\pi R^2 x dx$. This power is obviously split into two terms:
\begin{equation}
P=P_{ext} + P_{int} \, ,
\end{equation}
where $P_{ext}$ is the power transferred outside the waveguide, and $P_{int}$ is the power transferred inside.

The integrals mentioned above can be found in the literature. For convenience, we'll write out the corresponding formulas in a slightly modified form:
\begin{equation}
\int\limits_1^{\infty} K^2_1(at)tdt = \frac{1}{2}K^2_0(a) + 
\frac{1}{a}K_0(a)K_1(a) - \frac{1}{2}K^2_1(a) \, ,
\end{equation}
\begin{equation}
\int\limits_0^1 I^2_1(at)tdt = \frac{1}{2}I^2_1(a) + \frac{1}{a}I_0(a)I_1(a) - 
\frac{1}{2}I_0^2(a) \, .
\end{equation}
In accordance with this, the following is obtained:
\begin{equation}
P_{ext} =\frac{1}{2} I^2 W_{ext} \, ,
\end{equation}
\begin{equation}
P_{int} = \frac{1}{2}I^2 W_{int} \, ,
\end{equation}
where
\begin{equation}
\begin{array}{l}
\displaystyle
W_{ext} = \frac{2\pi R^2 \mu^2\xi W_0}{h^2B^2}
I^2_1\left(A\sqrt{\xi^2-\varepsilon\mu}\right) 
\left( \frac
{K^2_1\left(A\sqrt{\xi^2-1}\right)}
{(\xi^2-1)\tan^2\phi K^2_0\left(A\sqrt{\xi^2-1}\right)} + 1
 \right) \times \\ \\
\displaystyle \times
\left(\frac{1}{2}K^2_0(A\sqrt{\xi^2-1}) + 
\frac{1}{A\sqrt{\xi^2-1}}K_0(A\sqrt{\xi^2-1})K_1(A\sqrt{\xi^2-1})
 - \frac{1}{2}K^2_1(A\sqrt{\xi^2-1})\right) 
 \, ,
\end{array}
\end{equation}
\begin{equation}
\begin{array}{l}
\displaystyle

W_{int} = \frac{2\pi R^2 \mu\xi\ W_0}{h^2 B^2}\cdot K^2_1\left(A\sqrt{\xi^2-1}\right) 
\left(\frac {\varepsilon\mu I^2_1\left(A\sqrt{\xi^2-\varepsilon\mu}\right)}
{(\xi^2-\varepsilon\mu)\tan^2\phi I^2_0\left(A\sqrt{\xi^2-\varepsilon\mu}\right)}
 +  1 \right) \times \\ \\
\displaystyle \times
\left(
 \frac{1}{2}I^2_1\left(A\sqrt{\xi^2-\varepsilon\mu}\right) + 
 \frac{1}{A\sqrt{\xi^2-\varepsilon\mu}}
 I_0\left(A\sqrt{\xi^2-\varepsilon\mu}\right)I_1\left(A\sqrt{\xi^2-\varepsilon\mu}\right) - 
\frac{1}{2}I_0^2\left(A\sqrt{\xi^2-\varepsilon\mu}\right)
\right)
 \, .
\end{array}
\end{equation}

Both $P_{ext}$ and $P_{int}$ are proportional to the same square of the current amplitude. Therefore, the total power is also proportional to this square. The resulting formula is completely identical to the formula for the power transferred along a line with characteristic impedance.
\begin{equation}
W=W_{ext}+W_{int} \, .
\end{equation} 
So this sum is the wave impedance of the waveguide.

When considering radio engineering problems, a waveguide can be replaced by a line with such a characteristic impedance. It's important to remember that, unlike a conventional line, the characteristic impedance of such an equivalent line varies for different frequencies. The voltage $U$ in the wave traveling along such a line can also be introduced using the obvious formula
\begin{equation}
U = \pm W I \, ,
\end{equation}
where the plus sign should be taken for the incident wave, and the minus sign for the reflected wave. Naturally, the deceleration factor must be taken into account.

Figure 2 shows graphs of wave resistance as a function of wavelength in free space, for the same parameters as in Figure 1. For other parameters, this resistance can be calculated using the formulas given above.

\begin{figure}[h]
\begin{center}
\vspace{1cm}
\includegraphics[width=14cm, keepaspectratio]{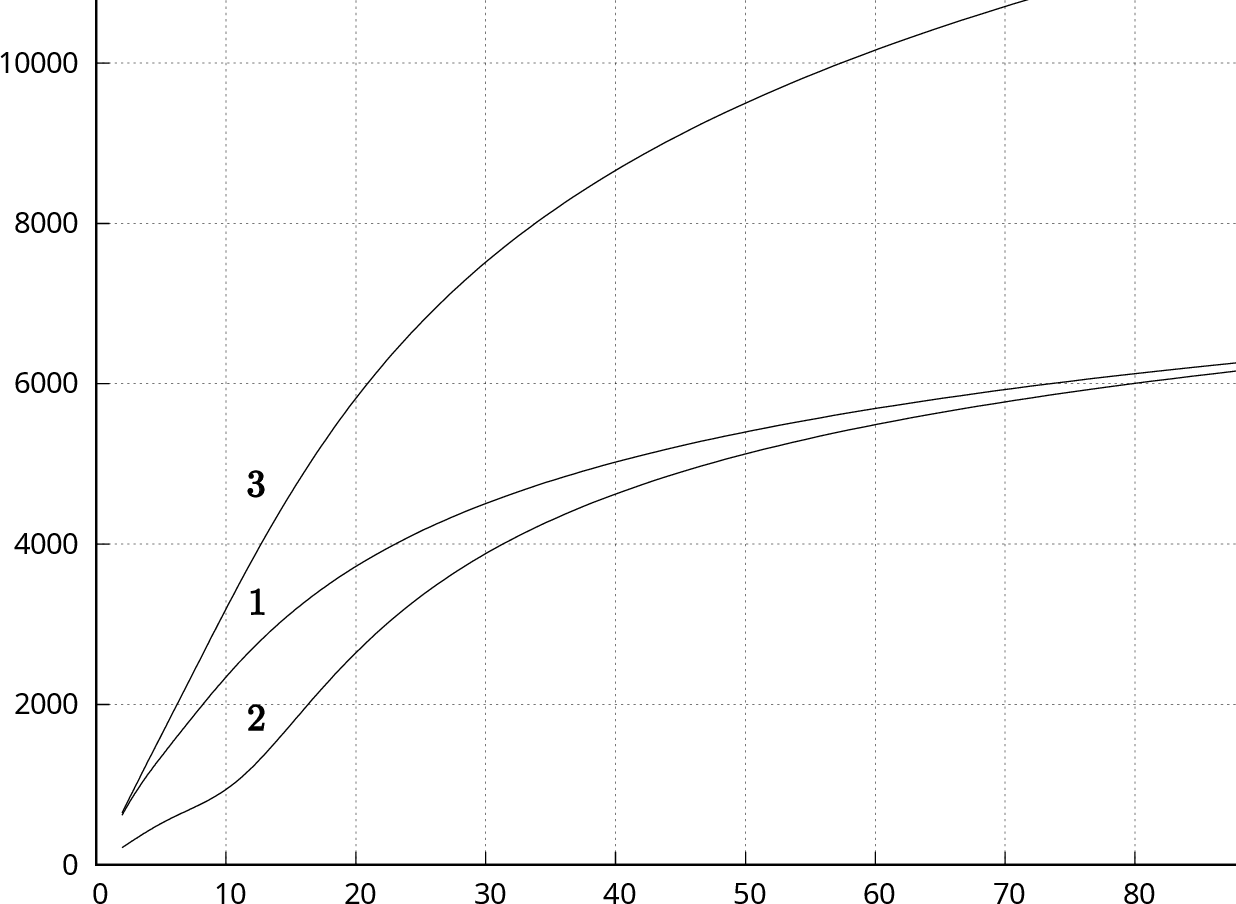}
\end{center}
\caption{Wave impedance $W$ depending on wavelength $\lambda$ for a waveguide with a diameter of 30 mm, with a winding pitch of 1 mm. 1: $\varepsilon=\mu=1$; 2: $\varepsilon=5$, $\mu=1$; 3: $\varepsilon=1$, $\mu=5$.}
\label{fig1}
\end{figure}

\clearpage

\section{Conclusion}

This paper examines the approximate theory of a helical waveguide for the case where the waveguide contains a dielectric with a certain permittivity and magnetic permeability. A dispersion equation is derived that determines the wave slowing coefficient. This equation is transcendental, so its analytical solution is not possible. However, its numerical solution by dividing a segment into halves presents no significant problem, as demonstrated in several specific examples. Closed analytical expressions are obtained for all components of the electromagnetic field. All these components are proportional to the current flowing through the winding. The power channeled by such a waveguide is considered. In this section, closed analytical formulas are also obtained; although rather cumbersome, they are quite suitable for practical calculations. In practical radio engineering problems, such a waveguide can be replaced by an equivalent long line with parameters that can be determined from the results of this paper. It is important to remember that these parameters, unlike those of a real line, depend on frequency.

The work was financed in accordance with the state assignment of the Omsk Scientific Center of the Siberian Branch of the Russian Academy of Sciences (state registration number of the project 125013101211-4).



\bibliographystyle{elsarticle-num}
\bibliography{The.bib}

\end{document}